\documentclass[aps,prb,twocolumn,showpacs]{revtex4}

\usepackage{graphicx}
\usepackage{amsmath}
\usepackage{amssymb}

\begin{document}

\title{Anomalous electron spectrum and its relation to peak structure of electron scattering rate in cuprate superconductors}

\author{Deheng Gao, Yingping Mou, and Shiping Feng}
\email{spfeng@bnu.edu.cn}

\affiliation{Department of Physics, Beijing Normal University, Beijing 100875, China}

\begin{abstract}
The recent discovery of a direct link between the sharp peak in the electron quasiparticle scattering rate of cuprate superconductors and the well-known peak-dip-hump structure in the electron quasiparticle excitation spectrum is calling for an explanation. Within the framework of the kinetic-energy driven superconducting mechanism, the complicated line-shape in the electron quasiparticle excitation spectrum of cuprate superconductors is investigated. It is shown that the interaction between electrons by the exchange of spin excitations generates a notable peak structure in the electron quasiparticle scattering rate around the antinodal and nodal regions. However, this peak structure disappears at the hot spots, which leads to that the striking peak-dip-hump structure is developed around the antinodal and nodal regions, and vanishes at the hot spots. The theory also confirms that the sharp peak observed in the electron quasiparticle scattering rate is directly responsible for the remarkable peak-dip-hump structure in the electron quasiparticle excitation spectrum of cuprate superconductors.
\end{abstract}

\maketitle

\section{Introduction}

The parent compound of cuprate superconductors is a Mott insulator, which occurs to be due to the strong electron correlation \cite{Anderson87,Phillips10}. As the charge carriers are introduced into this parent Mott insulator, a process called doping, the material evolves from an insulator to a superconductor \cite{Bednorz86,Kastner98}. This marked evolution of the electronic states indicates that the strong electron correlation in cuprate superconductors plays an essential role in the appearance of superconductivity, and also leads to the emergence of the unusual properties in both the superconducting (SC) and normal states \cite{Kastner98,Timusk99,Hufner08}. In conventional superconductors, as explained by the Bardeen-Cooper-Schrieffer theory \cite{Bardeen57,Schrieffer64}, superconductivity is caused by the interaction between electrons by the exchange of phonons. These exchanged phonons act like a bosonic glue to hold the electron pairs together. However, the mechanism of superconductivity in cuprate superconductors is still debated. Since the SC-state electron quasiparticle excitations determined by the electronic structure is closely related to the bosonic glue forming electron pairs \cite{Carbotte11}, the understanding of the nature of the electron quasiparticle excitations of cuprate superconductors is thought to be key to the understanding of the electron pairing mechanism.

Experimentally, by virtue of systematic studies using the angle-resolved photoemission spectroscopy (ARPES), the essential feature of the electron quasiparticle excitation spectrum of cuprate superconductors in the SC-state is well-established \cite{Damascelli03,Campuzano04,Zhou07}, where one of the most definite characteristics is the dramatic change in the spectral line-shape. The early ARPES experiments observed that around the antinodal region, a sharp electron quasiparticle excitation peak develops at the lowest binding energy corresponding to the SC gap, and is followed by a dip and then a hump in the higher energies, giving rise to a striking peak-dip-hump (PDH) structure in the electron quasiparticle excitation spectrum \cite{Dessau91,Hwu91,Randeria95,Saini97,Fedorov99,Lu01,Sato02,DLFeng02,Borisenko03}. In particular, this similar PDH structure has been observed in tunneling and Raman spectra \cite{Renner95,Fischer07,Devereaux07,Sakai13}. More importantly, the Raman scattering measurements found that the remarkable PDH structure is also developed around the nodal region \cite{Sakai13,Loret17,Jing17}. Therefore this well-known PDH structure now has been a hallmark of the spectral line-shape of the ARPES spectrum in cuprate superconductors. However, the recent ARPES experimental results \cite{DMou17} indicated a strong enhancement of the electron quasiparticle scattering rate around the antinodal region that manifests itself as a sharp peak in the imaginary part of the electron self-energy, and then this sharp peak is directly responsible for the famous PDH structure in the electron quasiparticle excitation spectrum \cite{DMou17}.

In spite of a general agreement on the importance of the PDH structure to superconductivity, the finial consensus on the physical origin of the PDH structure has not reached \cite{Carbotte11}. The earlier works \cite{Carbotte11} gave the main impetus for a phenomenological description of the electron quasiparticle excitations in terms of the strong interaction of electrons with a collective mode, which may be of fundamental relevance to the bosonic glue to hold the electron pairs together. In particular, a qualitative agreement is obtained in terms of the electron self-energy due to the interaction of electrons with a sharp spin resonance mode of the wave vector $(\pi,\pi)$ seen in the inelastic neutron scattering experiments \cite{Norman97,Eschrig00}. An important fact underlying these phenomenological analyses is the disappearance of the spin resonance at the SC transition temperature $T_{\rm c}$. However, it has been questioned that the spin resonance has too small a spectral weight \cite{Birgeneau89,Cheong91,Bourges00}, and it may be insufficient to produce superconductivity \cite{Kee02,Vojta11}. Thus it is rather difficult to obtain conclusive results. On the other hand, the further ARPES studies showed that the hump scales with the peak and persists above $T_{\rm c}$ in the pseudogap phase \cite{Campuzano99,Hashimoto15}, indicating that the striking PDH structure may be totally unrelated to superconductivity. These experimental results also show that the same interaction of electrons with the bosonic excitations that induces the SC-state in the particle-particle channel should also generate an obvious peak structure in the imaginary part of the electron self-energy in the particle-hole channel, and then this peak structure can evolve into the normal-state pseudogap phase \cite{Campuzano99,Hashimoto15}.

The strong electron correlation in cuprate superconductors comes from a large on-site repulsion between two electrons occupying the same site, which effectively translates into an elimination of the double electron occupancy. Apart from the numerical techniques, a powerful method to implement this elimination of the double electron occupancy is the so-called charge-spin separation slave-particle approach \cite{Feng93,Zhang93,Yu92,Kotliar88}, where the constrained electron is decoupled according its charge and spin degrees of freedom. However, a microscopic theory based on the charge-spin separation can not give a consistent description of the electron Fermi surface (EFS) and the related electron quasiparticle excitations in terms of the conventional charge-spin recombination \cite{Feng93,Zhang93}. In the recent work based on the kinetic-energy driven SC mechanism, we \cite{Feng15a} have developed a full charge-spin recombination scheme to fully recombine a charge carrier and a localized spin into an electron, where the electron self-energies in both the particle-particle and particle-hole channels are generated by the strong interaction between electrons by the exchange of spin excitations, and then the single-electron Green's function (then the electron self-energy) in the normal-state can produce a large EFS with an area satisfying Luttinger's theorem \cite{Feng16}. In this paper, we study the complicated line-shape in the electron quasiparticle excitation spectrum of cuprate superconductors along with this line \cite{Feng15a}. We show that the electron quasiparticle scattering rate arising from the electron self-energies exhibits a particularly obvious peak structure around the antinodal and nodal regions. However, this peak structure vanishes at hot spots, which cause a notable feature in the spectral line-shape generating the remarkable PDH structure in the electron quasiparticle excitation spectrum around the antinodal and nodal regions, in qualitative agreement with the experimental results \cite{Dessau91,Hwu91,Randeria95,Saini97,Fedorov99,Lu01,Sato02,DLFeng02,Borisenko03,Renner95,Fischer07,Devereaux07,Sakai13,Loret17,Jing17,DMou17}.
As a natural consequence of the disappearance of the peak structure in the electron quasiparticle scattering rate at the hot spots, the PDH structure in the electron quasiparticle excitation spectrum is absent from the hot spots.

The rest of this paper is organized as follows. The general formalism of the electron quasiparticle excitation spectrum of the $t$-$J$ model in the SC-state is presented in Sec. \ref{framework}, while the quantitative characteristics of the electron quasiparticle excitation spectrum of cuprate superconductors are discussed in Section \ref{line-shape}, where we show that there is one to one correspondence between the peak structure in the electron quasiparticle scattering rate of cuprate superconductors and the PDH structure in the electron quasiparticle excitation spectrum. In particular, this peak structure in the electron quasiparticle scattering rate and the related PDH structure in the electron quasiparticle excitation spectrum can be attributed to the emergence of the pseudogap. Finally, we give a summary in Sec. \ref{conclusions}.

\section{Formalism}\label{framework}

The energy and momentum dependence of the ARPES spectrum can be described by \cite{Damascelli03,Campuzano04,Zhou07},
\begin{eqnarray}\label{ARPES}
I({\bf k},\omega)=|M({\bf k},\omega)|^{2}n_{\rm F}(\omega)A({\bf k},\omega),
\end{eqnarray}
where $M({\bf k},\omega)$ is a dipole matrix element that depends on the initial and final electronic states, incident photon energy, and polarization. However, as a qualitative discussion in this paper, the magnitude of the dipole matrix element $M({\bf k},\omega)$ has been rescaled to the unit. $n_{\rm F}(\omega)$ is the fermion distribution, while the electron spectral function $A({\bf k},\omega)$ in the SC-state is related directly to the imaginary part of the single-electron diagonal Green's function $G({\bf k},\omega)$ as $A({\bf k},\omega)=-2{\rm Im}G({\bf k}, \omega)$. This ARPES spectrum in Eq. (\ref{ARPES}) yields a detailed picture of the energy and momentum dependence of the electronic structure of the occupied states below EFS.

Very soon after the discovery of superconductivity in cuprate superconductors \cite{Bednorz86}, it has been argued that the essential physics of cuprate superconductors is captured by the $t$-$J$ model on a square lattice \cite{Anderson87}. Since then many elaborations based on this $t$-$J$ model followed \cite{Lee06,Edegger07}. In particular, it has been suggested that the spin excitation in cuprate superconductors, which is a generic consequence of the strong electron correlation, can mediate the electron pairing \cite{Lee06,Edegger07,Monthoux07}. On the other hand, the combined inelastic neutron scattering and resonant inelastic X-ray scattering experimental data have identified the spin excitations with high intensity over a large part of moment space, and shown that the spin excitations exist across the entire range of the SC dome, and with sufficient intensity to mediate superconductivity in cuprate superconductors \cite{Fujita12,Dean15}. For the understanding of the SC-state properties of cuprate superconductors, we \cite{Feng0306,Feng12,Feng15} have developed a kinetic-energy driven SC mechanism based on the $t$-$J$ model in the charge-spin separation fermion-spin representation, where the interaction between charge carriers and spins directly from the kinetic energy by the exchange of spin excitations generates the charge-carrier pairing state in the particle-particle channel and the charge-carrier pseudogap state in the particle-hole channel, while the electron pairs originated from the charge-carrier pairing state are due to the charge-spin recombination \cite{Feng15a}, and their condensation reveals the SC ground-state. In the following discussions, we reproduce only the main details in the calculation of the single-electron Green's function of the $t$-$J$ model. In Ref. \onlinecite{Feng15a}, the single-electron diagonal and off-diagonal Green's functions of the $t$-$J$ model in the SC-sate have been obtained in terms of the full charge-spin recombination scheme as,
\begin{widetext}
\begin{subequations}\label{EGFS}
\begin{eqnarray}
G({\bf k},\omega)&=&{1\over \omega-\varepsilon_{\bf k}-\Sigma_{1}({\bf k},\omega)-[\Sigma_{2}({\bf k},\omega)]^{2}/[\omega+\varepsilon_{\bf k}+ \Sigma_{1}({\bf k},-\omega)]}, \label{DEGF}\\
\Im^{\dagger}({\bf k},\omega)&=&-{\Sigma_{2}({\bf k},\omega)\over [\omega-\varepsilon_{\bf k}-\Sigma_{1}({\bf k},\omega)][\omega+\varepsilon_{\bf k}+ \Sigma_{1}({\bf k},-\omega)]-[\Sigma_{2}({\bf k},\omega)]^{2}},\label{ODEGF}
\end{eqnarray}
\end{subequations}
\end{widetext}
where the bare electron excitation spectrum $\varepsilon_{\bf k}=-Zt\gamma_{\bf k}+Zt'\gamma_{\bf k}'+\mu$, with $\gamma_{\bf k}=({\rm cos}k_{x}+{\rm cos}k_{y})/2$, $\gamma_{\bf k}'= {\rm cos} k_{x}{\rm cos}k_{y}$, the nearest-neighbor (NN) hopping amplitude $t$, the next NN hopping amplitude $t'$, and $Z$ is the number of the NN or next NN sites on a square lattice, while the electron self-energies $\Sigma_{1}({\bf k},\omega)$ in the particle-hole channel and $\Sigma_{2}({\bf k},\omega)$ in the particle-particle channel have been evaluated in terms of the full charge-spin recombination, and are given explicitly in Ref. \onlinecite{Feng15a}. As in the case of the previous studies \cite{Feng15a}, the parameters are chosen as $t/J=2.5$ and $t'/t=0.3$, where $J$ is the NN spin-spin antiferromagnetic exchange in the $t$-$J$ model. It should be emphasized that in the full charge-spin recombination scheme \cite{Feng15a}, the coupling form between the electron quasiparticles and spin excitations is the same as that between the charge-carrier quasiparticles and spin excitations, indicating that both the electron self-energies $\Sigma_{1}({\bf k},\omega)$ in the particle-hole channel and $\Sigma_{2}({\bf k},\omega)$ in the particle-particle channel are induced by the strong interaction between electrons by the exchange of spin excitations.

In the strong coupling formalism \cite{Eliashberg60,Mahan81}, both the energy and momentum dependence of the pairing force and electron pair order parameter have been incorporated into the electron self-energy $\Sigma_{2}({\bf k},\omega)$ in the particle-particle channel. In this sense, $\Sigma_{2}({\bf k},\omega)$ represents the energy and momentum dependence of the electron pair gap $\bar{\Delta}_{\rm s}({\bf k},\omega)= \Sigma_{2}({\bf k},\omega)$, where following the common practice, the imaginary part of $\Sigma_{2}({\bf k},\omega)$ has been neglected. On the other hand, the electron self-energy $\Sigma_{1}({\bf k},\omega)$ in the particle-hole channel can be separated into two parts: $\Sigma_{1}({\bf k},\omega) ={\rm Re}\Sigma_{1}({\bf k},\omega)+i{\rm Im}\Sigma_{1}({\bf k}, \omega)$, with ${\rm Re} \Sigma_{1}({\bf k},\omega)$ and ${\rm Im}\Sigma_{1}({\bf k},\omega)$ that are, respectively, the corresponding real and imaginary parts of $\Sigma_{1}({\bf k},\omega)$. With the above electron diagonal Green's function (\ref{DEGF}), the electron spectral function $A({\bf k},\omega)$ in the SC-state now can be obtained explicitly as,
\begin{eqnarray}\label{ESF}
A({\bf k},\omega)={2\Gamma({\bf k},\omega)\over [\omega-\varepsilon_{\bf k}-{\rm Re}\bar{\Sigma}({\bf k},\omega)]^{2}+\Gamma^{2}({\bf k},\omega)},
\end{eqnarray}
and then the electron quasiparticle excitation spectrum $I({\bf k},\omega)$ in Eq. (\ref{ARPES}) can be measurable via the ARPES technique \cite{Damascelli03,Campuzano04,Zhou07}, where the electron quasiparticle scattering rate $\Gamma({\bf k},\omega)$ and the real part of the modified electron self-energy ${\rm Re}\bar{\Sigma}({\bf k},\omega)$ can be expressed as,
\begin{widetext}
\begin{subequations}\label{MESE}
\begin{eqnarray}
\Gamma({\bf k},\omega)&=&\left | {\rm Im}\Sigma_{1}({\bf k},\omega)-{\bar{\Delta}^{2}_{\rm s}({\bf k},\omega){\rm Im}\Sigma_{1}({\bf k},-\omega)\over [\omega +\varepsilon_{\bf k}+{\rm Re}\Sigma_{1}({\bf k},-\omega)]^{2}+[{\rm Im}\Sigma_{1}({\bf k},-\omega)]^{2}}\right |, \label{EQDSR}\\
{\rm Re}\bar{\Sigma}({\bf k},\omega)&=&{\rm Re}\Sigma_{1}({\bf k},\omega)+{\bar{\Delta}^{2}_{\rm s}({\bf k},\omega)[\omega+\varepsilon_{\bf k}+{\rm Re}\Sigma_{1}({\bf k},-\omega)]\over [\omega+\varepsilon_{\bf k}+{\rm Re}\Sigma_{1}({\bf k},-\omega)]^{2}+[{\rm Im}\Sigma_{1}({\bf k},-\omega)]^{2}}, \label{MRESE}
\end{eqnarray}
\end{subequations}
\end{widetext}
respectively.

\section{Correlation between peak in electron scattering rate and peak-dip-hump structure in electron spectrum} \label{line-shape}

In the SC-state, the contribution to the electron quasiparticle excitation spectrum comes from two typical excitations: the electron-hole and the electron pair excitations, which leads to a rather complicated form in the electron spectral function (\ref{ESF}), where the real part of the modified electron self-energy ${\rm Re}\bar{\Sigma}({\bf k},\omega)$ in Eq. (\ref{MRESE}) reduces electron quasiparticle dispersion, while the lifetime of the electron quasiparticle excitation is determined by the electron quasiparticle scattering rate $\Gamma({\bf k},\omega)$.

\begin{figure*}[t!]
\centering
\includegraphics[scale=0.35]{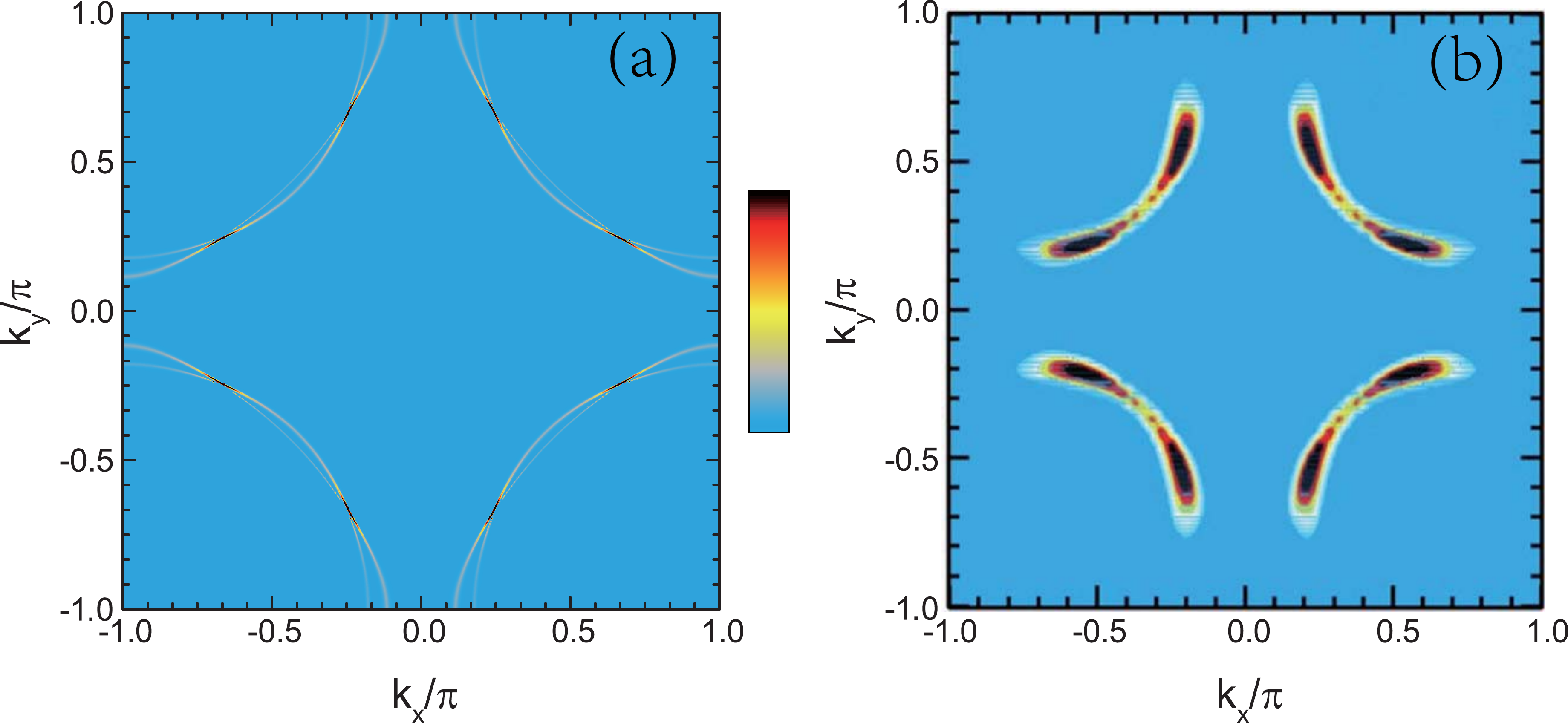}
\caption{(Color online) (a) The map of the spectral intensity of the electron quasiparticle excitation spectrum in the $[k_{x},k_{y}]$ plane for $\omega=-24$ meV at $\delta=0.15$ with $T=0.002J$ for $t/J=2.5$ and $t'/t=0.3$. (b) The corresponding experimental result of the optimally doped Bi$_{2}$Sr$_{2}$CaCu$_{2}$O$_{8+\delta}$ for binding energy $\omega=-24$ meV taken from Ref. \onlinecite{Chatterjee06}.
\label{spectrum-map}}
\end{figure*}

Firstly, we plot a map of the spectral intensity of the electron quasiparticle excitation spectrum $I({\bf k},\omega)$ in the Brillouin zone (BZ) for the binding energy $\omega=-24$ meV at doping $\delta=0.15$ with temperature $T=0.002J$ in Fig. \ref{spectrum-map}a. For a comparison, the corresponding experimental result \cite{Chatterjee06} of the ARPES spectral intensity map observed from the optimally doped Bi$_{2}$Sr$_{2}$CaCu$_{2}$O$_{8+\delta}$ for the binding energy $\omega=-24$ meV in the SC-state is also shown in Fig. \ref{spectrum-map}(b). It is shown clearly that the corresponding ARPES experimental result \cite{Chatterjee06} is qualitatively reproduced. In particular, as in the previous case of the study of the nature of EFS [the map of the spectral intensity of the electron quasiparticle excitation spectrum $I({\bf k},\omega)$ at zero energy ($\omega=0$)] in the normal-state \cite{Yang08,Chang08,Meng09,Yang11,Zhao17}, two characteristic features emerge: (i) the coexistence of the disconnected segments and pockets around the nodal region, and (ii) the highest intensity points located at the tips of the disconnected segments, which in this case coincide with the hot spots on the constant energy contours. These hot spots connected by the scattering wave vector contribute effectively to the electron quasiparticle scattering process. In particular, the electron quasiparticle scattering between two hot spots on the straight disconnected segments causes the charge ordering instability \cite{Feng16,Comin15,Wu11,Ghiringhelli12,Comin14,Campi15,Mou17}. Furthermore, we \cite{Gao17} have also discussed the autocorrelation ${\bar C}({\bf q},\omega)=(1/N)\sum_{\bf k}I({\bf k}+{\bf q},\omega)I({\bf k},\omega)$ of the electron quasiparticle excitation spectral intensities, and found that the spots in ${\bar C}({\bf q},\omega)$ are directly correlated with those wave vectors ${\bf q}_{i}$ that connect the tips of the disconnected segments in Fig. \ref{spectrum-map}a, in qualitative agreement with the ARPES experimental data \cite{Chatterjee06}. More importantly, these ${\bf q}_{i}$ wave vectors are also qualitatively consistent with those observed from the Fourier transform scanning tunneling spectroscopy experiments \cite{Hoffman02,McElroy03}, indicating that the {\it octet} model that has been used to explain the Fourier transform scanning tunneling spectroscopy experimental data can give a consistent description of the regions of the highest joint density of states.

\begin{figure*}[t!]
\centering
\includegraphics[scale=0.55]{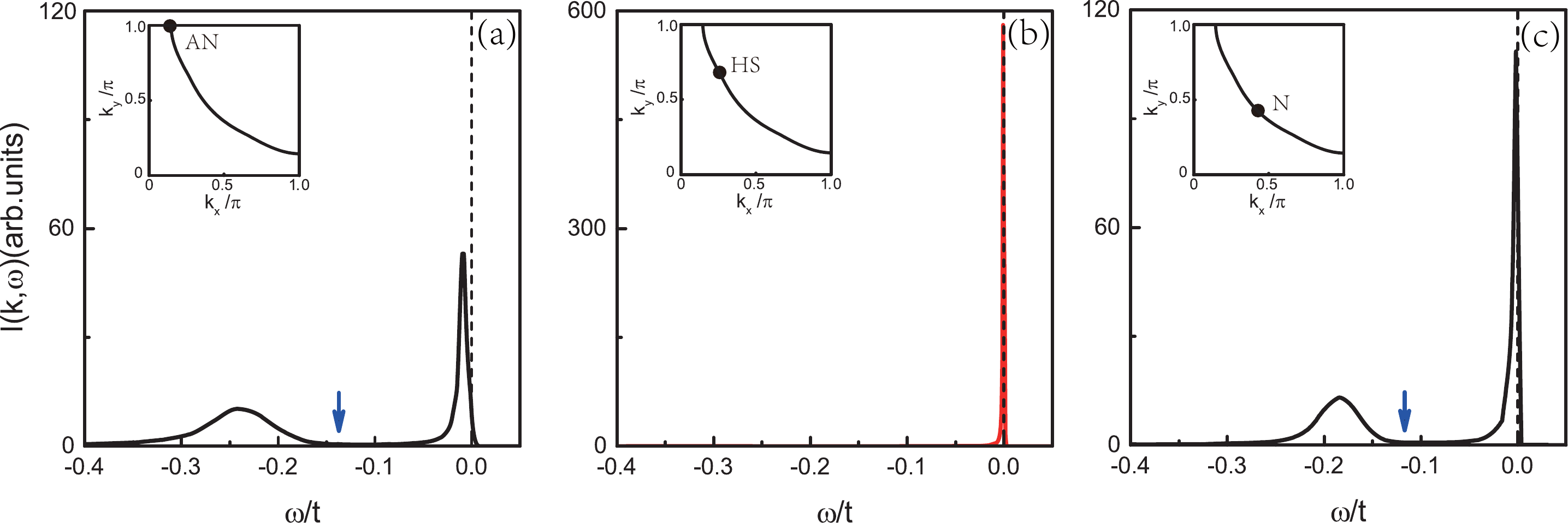}
\caption{(Color online) The electron quasiparticle excitation spectrum on the electron Fermi surface as a function of energy around (a) the antinode, (b) the hot spot, and (c) the node at $\delta=0.15$ with $T=0.002J$ for $t/J=2.5$ and $t'/t=0.3$, where the blue arrow indicates the position of the dip, and AN, HS, and N in the insets denote the antinode, hot spot, and node, respectively.
\label{PDH}}
\end{figure*}

\begin{figure}[h!]
\centering
\includegraphics[scale=0.4]{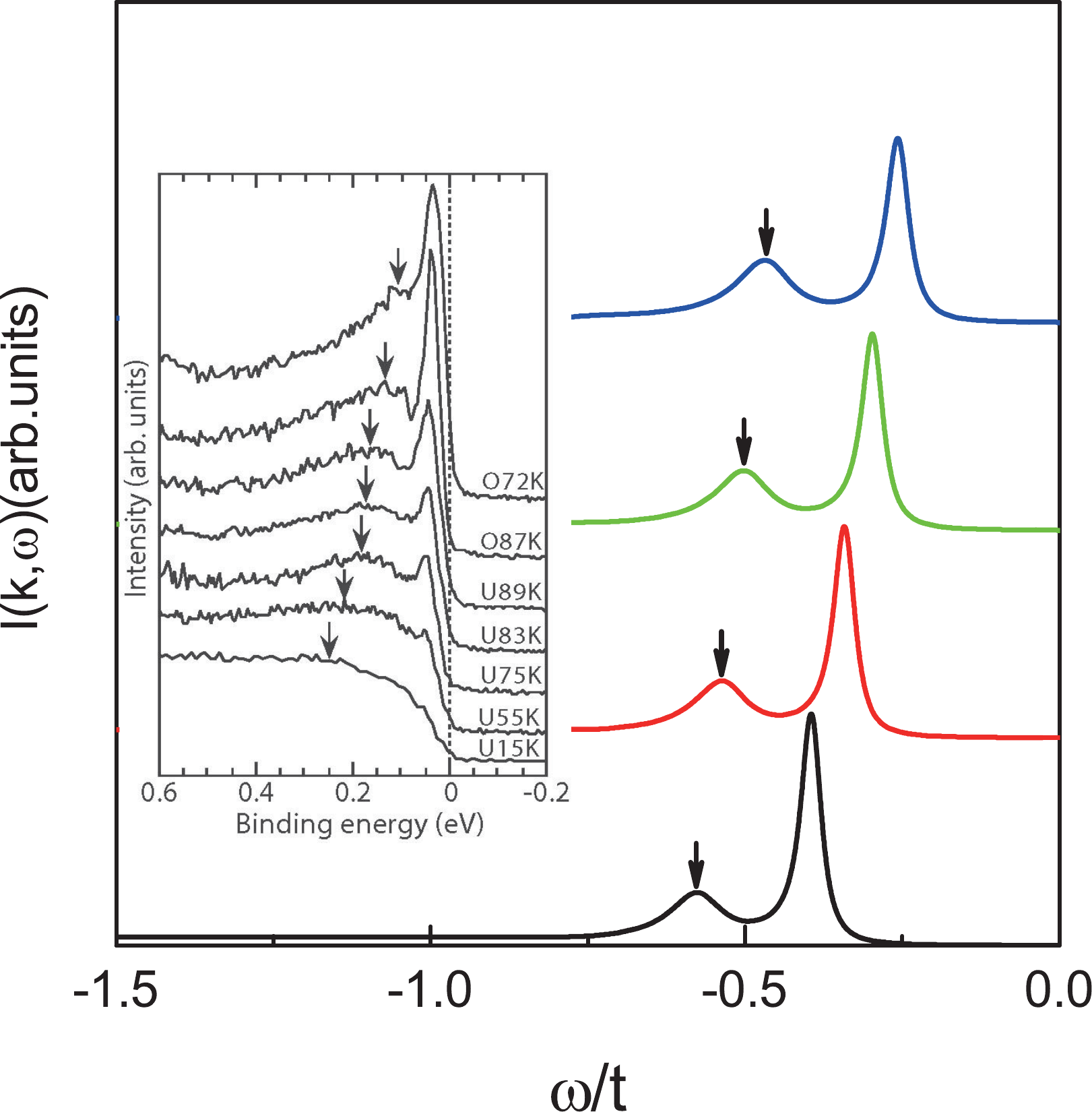}
\caption{(Color online) The electron quasiparticle excitation spectrum at $(\pi,0)$ point of the Brillouin zone as a function of energy with $T=0.002J$ in $\delta=0.09$ (black line), $\delta=0.12$ (red line), $\delta=0.15$ (green line), and $\delta=0.18$ (blue line) for $t/J=2.5$ and $t'/t=0.3$. Inset: the corresponding experimental result of Bi$_{2}$Sr$_{2}$CaCu$_{2}$O$_{8+\delta}$ taken from Ref. \onlinecite{Campuzano99}.
\label{PDH-doping}}
\end{figure}

With the help of the above map of the spectral intensity of the electron quasiparticle excitation spectrum in Fig. \ref{spectrum-map}, we now turn to discuss the complicated line-shape in the electron quasiparticle excitation spectrum of cuprate superconductors. We have made a series of calculations for $I({\bf k},\omega)$ along with EFS from the antinode to node, and the results of $I({\bf k},\omega)$ as a function of energy around (a) the antinode, (b) the hot spot, and (c) the node at $\delta=0.15$ with $T=0.002J$ are plotted in Fig. \ref{PDH}. At the antinode, the electron quasiparticle excitation spectrum consists of two separate peaks: a very sharp low-energy peak and a relatively broad high-energy peak, which are corresponding to the SC quasiparticle excitation and hump, respectively. Between these two peaks is a dip, which corresponds to the intensity depletion region. The total contributions for the electron quasiparticle excitation spectrum therefore give rise to the PDH structure (see Fig. \ref{PDH}a), where the position of the sharp low-energy peak deviates from EFS is due to the opening of the d-wave type SC gap. However, the positions of the peak, dip, and hump are momentum dependent. In particular, the position of the hump appreciably shifts towards the low-energy peak as one moves away from the antinode to the hot spot, and eventually this hump disappears at the hot spot, leading to the absence of the PDH structure at the hot spots (see Fig. \ref{PDH}b). However, this PDH structure is gradually developed again as one moves away from the hot spot to the node, and then the PDH structure appears around the nodal region (see Fig. \ref{PDH}c). These results of the PDH structure in the electron quasiparticle excitation spectrum around the antinodal and nodal regions are qualitatively consistent with the experimental observations on cuprate superconductors in the SC-state \cite{Dessau91,Hwu91,Randeria95,Saini97,Fedorov99,Lu01,Sato02,DLFeng02,Borisenko03,Renner95,Fischer07,Devereaux07,Sakai13,Loret17,Jing17,DMou17}. Moreover, as a natural consequence of the doped Mott insulators, this PDH structure is also doping dependent. To see the evolution of the PDH structure with doping clearly, the results of $I({\bf k},\omega)$ as a function of energy at $(\pi,0)$ point of BZ with $T=0.002J$ for $\delta=0.09$ (black line), $\delta=0.12$ (red line), $\delta=0.15$ (green line), and $\delta=0.18$ (blue line) are plotted in Fig. \ref{PDH-doping} in comparison with the corresponding experimental data \cite{Campuzano99} of Bi$_{2}$Sr$_{2}$CaCu$_{2}$O$_{8+\delta}$ (inset). Apparently, the positions of both the hump and low-energy peak move to higher energies with the decrease of doping, which are also in qualitative agreement with the corresponding ARPES experimental results \cite{Campuzano99,Harris96}, where the similar doping dependence of the positions of the hump and low-energy peak have been observed on cuprate superconductors in the SC-state.

A natural question is why the PDH structure in the electron quasiparticle excitation spectrum of cuprate superconductors can be described qualitatively within the framework of the kinetic-energy driven superconductivity. The reason is that the presence of the spin excitations in cuprate superconductors has a well-pronounced effect on both the real and imaginary parts of the electron self-energy. This can be understood from the electron quasiparticle scattering rate $\Gamma({\bf k},\omega)$ in Eq. (\ref{EQDSR}) and the real part of the modified electron self-energy ${\rm Re} \bar{\Sigma}({\bf k},\omega)$ in Eq. (\ref{MRESE}). The electron quasiparticle excitation spectrum $I({\bf k}, \omega)$ in Eq. (\ref{ARPES}) [then the electron spectral function $A({\bf k}, \omega)$ in Eq. (\ref{ESF})] exhibits a peak when the incoming photon energy $\omega$ is equal to the electron quasiparticle excitation energy $E({\bf k})$, i.e.,
\begin{eqnarray}\label{band}
E_{\bf k}-\varepsilon_{\bf k}-{\rm Re}\bar{\Sigma}({\bf k},E_{\bf k})=0,
\end{eqnarray}
and then the lifetime of the electron quasiparticle excitation at the energy $\omega$ is determined by the inverse of the electron quasiparticle scattering rate $\Gamma({\bf k},\omega)$.

\begin{figure*}[t!]
\centering
\includegraphics[scale=0.385]{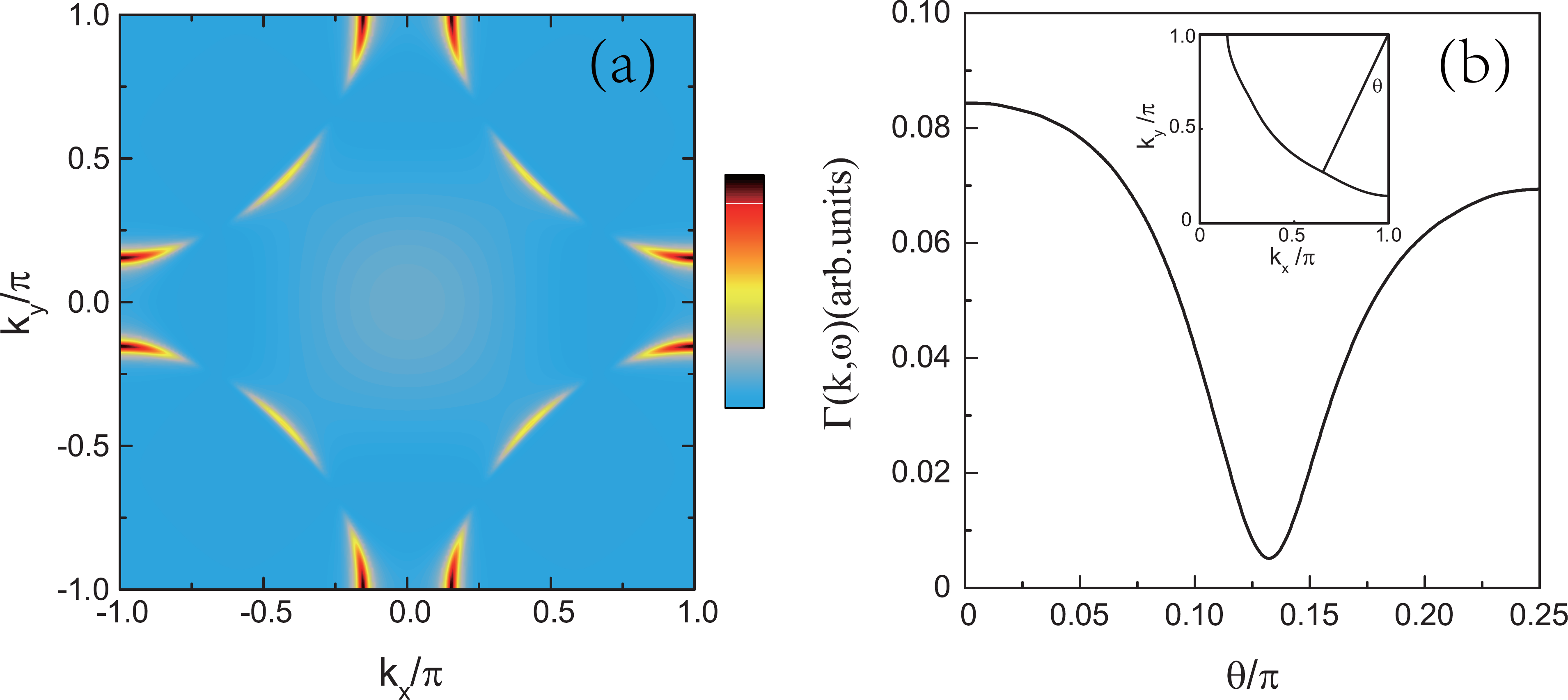}
\caption{(Color online) (a) The map of the intensity of the electron quasiparticle scattering rate and (b) the angular dependence of the the electron quasiparticle scattering rate on the constant energy contour shown in Fig. \ref{spectrum-map}a for $\omega=-24$ meV at $\delta=0.15$ with $T=0.002J$ for $t/J=2.5$ and $t'/t=0.3$. \label{scattering-rate-1}}
\end{figure*}

On the one hand, for a given energy $\omega$, $\Gamma({\bf k},\omega)$ varies strongly with momentum. To reveal this highly anisotropic $\Gamma({\bf k},\omega)$ in momentum space clearly, we plot (a) the map of the intensity of $\Gamma({\bf k},\omega)$ in BZ and (b) the angular dependence of $\Gamma({\bf k}, \omega)$ on the constant energy contour shown in Fig. \ref{spectrum-map}a for $\omega=-24$ meV at $\delta=0.15$ with $T=0.002J$ in Fig. \ref{scattering-rate-1}, where the actual minimum of $\Gamma({\bf k},\omega)$ does not appear around the node, but locates exactly at the hot spots. However, $\Gamma({\bf k},\omega)$ still exhibits the largest value around the antinode, and then it decreases with the move of the momentum away from the antinode. In particular, the magnitude of $\Gamma({\bf k},\omega)$ around the node is smaller than that around the antinode. This special momentum dependence of $\Gamma({\bf k},\omega)$ therefore suppresses heavily the low-energy spectral weight of the electron quasiparticle excitation spectrum around the antinodal region, but has a more modest effect on the spectral weight around the nodal region. In this case, the tips of these disconnected segments on the constant energy contours converge on the hot spots to form the closed pocket, generating a coexistence of the disconnected segments and pockets as shown in Fig. \ref{spectrum-map}a.

\begin{figure*}[t!]
\centering
\includegraphics[scale=0.585]{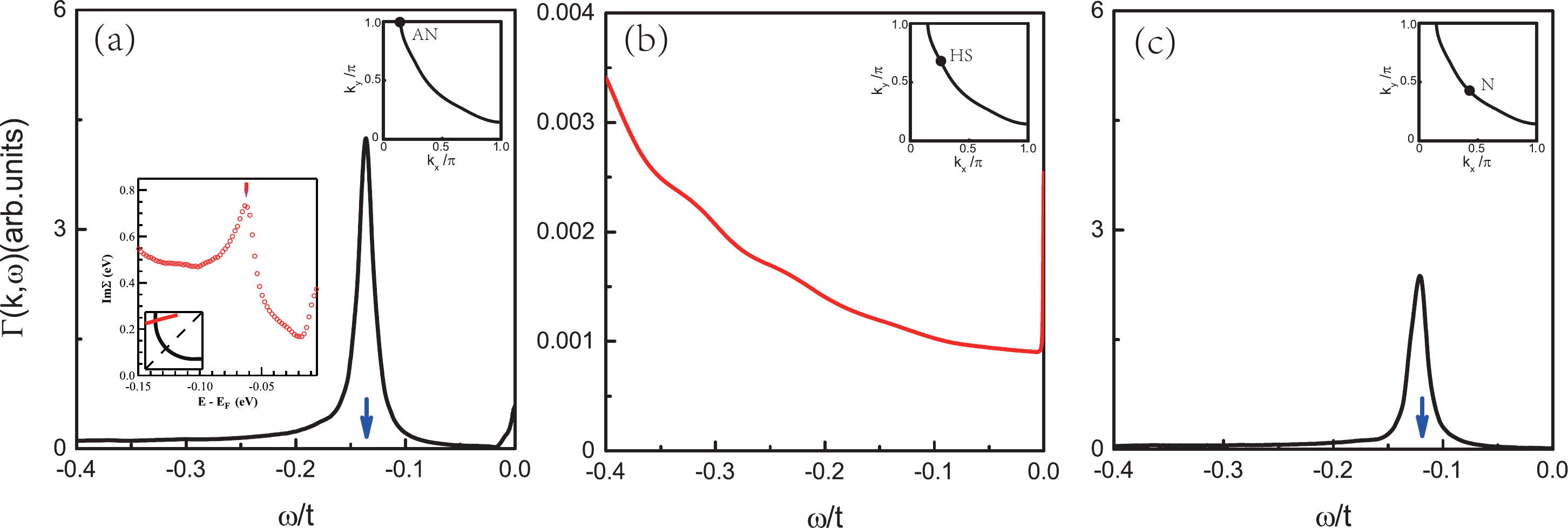}
\caption{(Color online) The electron quasiparticle scattering rate around (a) the antinode, (b) the hot spot, and (c) the node as a function of energy at $\delta=0.15$ with $T=0.002J$ for $t/J=2.5$ and $t'/t=0.3$, where the blue arrow indicates the position of the peak. Inset in (a): the corresponding experimental result of Bi$_{2}$Sr$_{2}$CaCu$_{2}$O$_{8+\delta}$ taken from Ref. \onlinecite{DMou17}. \label{scattering-rate-2}}
\end{figure*}

\begin{figure*}[t!]
\centering
\includegraphics[scale=0.585]{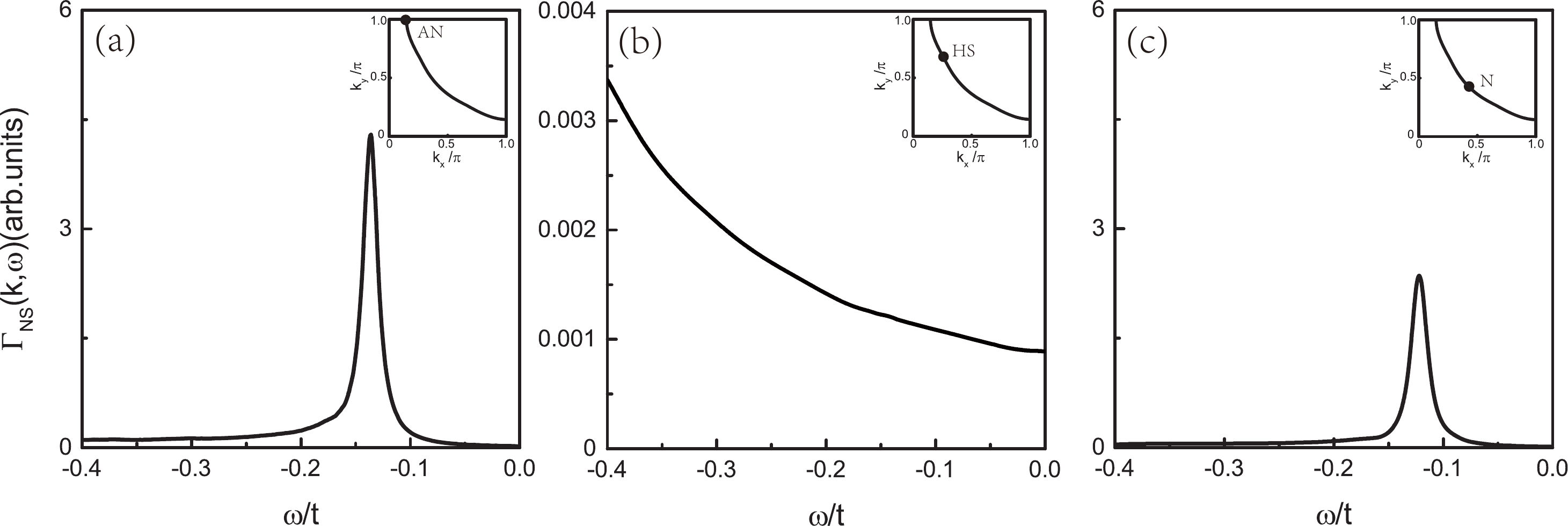}
\caption{The normal-state electron quasiparticle scattering rate around (a) the antinode, (b) the hot spot, and (c) the node as a function of energy at $\delta=0.15$ for $t/J=2.5$ and $t'/t=0.3$. \label{scattering-rate-3}}
\end{figure*}

On the other hand, for a given momentum ${\bf k}$, $\Gamma({\bf k},\omega)$ also varies strongly with energy. To see this point clearly, we plot $\Gamma({\bf k},\omega)$ as a function of energy around (a) the antinode, (b) the hot spot, and (c) the node at $\delta=0.15$ with $T=0.002J$ in Fig. \ref{scattering-rate-2}. For comparison, the corresponding experimental result \cite{DMou17} of the energy dependence of the electron quasiparticle scattering rate around the antinodal region found in the optimally doped Bi$_{2}$Sr$_{2}$CaCu$_{2}$O$_{8+\delta}$ is also shown in Fig. \ref{scattering-rate-2}a (inset). The result in Fig. \ref{scattering-rate-2}a shows clearly that around the antinodal region, a well-pronounced peak structure appears, where $\Gamma({\bf k},\omega)$ reaches a sharp peak at a binding energy of $-0.136t$, and then it decreases rapidly in both the low-energy and high-energy regimes. In particular, the position of the sharp peak is just corresponding to the position of the dip in the PDH structure in the electron quasiparticle excitation spectrum shown in Fig. \ref{PDH}a, and therefore the peak structure in $\Gamma({\bf k},\omega)$ induces an intensity depletion in $I({\bf k},\omega)$ around the dip. Moreover, using a reasonably estimative value of $J\sim 150$ meV, the anticipated position of the sharp peak at the binding energy of $-0.136t=-51$ meV in the optimal doping is not too far from the peak position at the binding energy of -62 meV observed \cite{DMou17} in the optimally doped Bi$_{2}$Sr$_{2}$CaCu$_{2}$O$_{8+\delta}$. Furthermore, as a comparison, we have also calculated the imaginary part of the electron self-energy ${\rm Im}\Sigma_{1} ({\bf k},\omega)$ as a function of energy for the same set of parameters as in Fig. \ref{scattering-rate-2}, and the result of ${\rm Im}\Sigma_{1} ({\bf k},\omega)$ at the antinode is almost the same as that of $\Gamma({\bf k},\omega)$ shown in Fig. \ref{scattering-rate-2}a except for the upturn of $\Gamma({\bf k},\omega)$ near EFS, where the upturn of ${\rm Im}\Sigma_{1} ({\bf k},\omega)$ near EFS is absent. These results therefore confirm that the upturn of $\Gamma({\bf k},\omega)$ near EFS is caused by the SC gap opening, and vanishes in the normal-state, in good agreement with the experimental data \cite{DMou17}. However, the sharp peak in $\Gamma({\bf k},\omega)$ is gradually suppressed as one moves away from the antinodal region, and then the peak structure vanishes eventually at the hot spots (see Fig. \ref{scattering-rate-2}b). Moreover, the peak structure in $\Gamma({\bf k},\omega)$ is gradually developed again as one moves away from the hot spot, and then the peak structure emerges around the nodal region, although the weight of the peak is much smaller than that around the antinodal region (see Fig. \ref{scattering-rate-2}c). This special energy and momentum dependence of the peak structure in $\Gamma({\bf k}, \omega)$ therefore leads to that the striking PDH structure in the electron quasiparticle excitation spectrum is developed around the antinodal and nodal regions, and disappears at the hot spots.

It should be noted that in the recent ARPES measurements \cite{DMou17}, the well-pronounced peak structure in the electron quasiparticle scattering rate (then the PDH structure in the electron quasiparticle excitation spectrum) emerges mainly around the antinodal region, which is consistent with the early ARPES experimental results \cite{Dessau91,Hwu91,Randeria95,Saini97,Fedorov99,Lu01,Sato02,DLFeng02,Borisenko03}. However, in a clear contrast to the early Raman scattering measurements results \cite{Sakai13,Loret17,Jing17}, the weak peak structure in the electron quasiparticle scattering rate around the nodal region was not observed \cite{DMou17}. On the other hand, our present theoretical results of the well-pronounced peak structure in the electron quasiparticle scattering rate around the antinodal region are well consistent with these observed in the recent ARPES experiments \cite{DMou17}, while the results of the electron quasiparticle scattering rate around the nodal region and the related constant energy contours at zero and finite energies are in qualitative agreement with the early ARPES experimental data \cite{Chatterjee06,Comin14} and Raman scattering measurements results \cite{Sakai13,Loret17}. However, the theory also predicts that the peak structure in the electron quasiparticle scattering rate is absent from the hot-spot directions, which should be verified by future experiments.

As a complement of the above analysis of the PDH structure in the SC-state, we now discuss the complicated line-shape of the electron quasiparticle excitation spectrum in the normal-state. In the normal-state [the SC gap $\bar{\Delta}_{\rm s}({\bf k},\omega)=0$], the SC-state electron quasiparticle scattering rate $\Gamma({\bf k},\omega)$ in Eq. (\ref{EQDSR}) is reduced as the normal-state electron quasiparticle scattering rate $\Gamma_{\rm NS}({\bf k},\omega)=|{\rm Im}\Sigma_{1}({\bf k}, \omega)|$. In Fig. \ref{scattering-rate-3}, we plot $\Gamma_{\rm NS}({\bf k},\omega)$ as a function of energy around (a) the antinode, (b) the hot spot, and (c) the node at $\delta=0.15$. Comparing it with Fig. \ref{scattering-rate-2} for the same set of parameters except for $\bar{\Delta}_{\rm s}({\bf k},\omega)=0$, we see that although the upturn of $\Gamma_{\rm NS}({\bf k},\omega)$ near EFS is absent, the main feature of the peak structure in the SC-state electron quasiparticle scattering rate persists into the normal-state. This peak structure in $\Gamma_{\rm NS}({\bf k},\omega)$ therefore leads to the similar energy and momentum dependence of the PDH structure in the electron quasiparticle excitation spectrum of cuprate superconductors in the normal-state \cite{Zhao17}, also in qualitative agreement with the experimental results \cite{Campuzano99,Hashimoto15}.

Within the framework of the kinetic-energy driven SC mechanism, the same electron interaction mediated by spin excitations that generates the SC-state in the particle-particle channel also induces the pseudogap state in the particle-hole channel \cite{Feng15a,Feng12,Feng15}. This follows a fact that the electron self-energy $\Sigma_{1}({\bf k},\omega)$ in the particle-particle channel can be also rewritten approximately as $\Sigma_{1}({\bf k},\omega) \approx [\bar{\Delta}_{\rm PG}({\bf k})]^{2}/[\omega+\varepsilon_{0{\bf k}}]$, where $\bar{\Delta}_{\rm PG}({\bf k})$ and $\varepsilon_{0{\bf k}}$ are the pseudogap and the energy spectrum, respectively, and have been given explicitly in Ref. \onlinecite{Feng15a}. In this case, the corresponding imaginary part of $\Sigma_{1}({\bf k},\omega)$ can be expressed in terms of the pseudogap as ${\rm Im}\Sigma_{1}({\bf k}, \omega)\approx -\pi [\bar{\Delta}_{\rm PG}({\bf k})]^{2}\delta(\omega+\varepsilon_{0{\bf k}})$, reflecting an intimate relation between the electron quasiparticle scattering rate and pseudogap \cite{Hashimoto15}. This pseudogap as a competing order persists up to the pseudogap crossover temperature $T^{*}$, and coexists with superconductivity below $T_{\rm c}$ \cite{Feng15a,Feng12,Feng15}. This pseudogap (then the electron quasiparticle scattering rate) generates the PDH structure in both the SC- and normal-states. In other words, the well-known PDH structure in the electron quasiparticle excitation spectrum can be attributed to the emergence of the pseudogap. This is also why the striking PDH structure is totally unrelated to superconductivity \cite{Campuzano99,Hashimoto15}. Furthermore, from the electron spectral function $A({\bf k},\omega)$ in Eq. (\ref{ESF}) and the related electron quasiparticle scattering rate $\Gamma({\bf k},\omega)$ in Eq. (\ref{EQDSR}) and real part of the modified electron self-energy ${\rm Re}\bar{\Sigma}({\bf k},\omega)$ in Eq. (\ref{MRESE}), it can be found that the positions of the hump and low-energy peak in the electron quasiparticle excitation spectrum are determined by both the pseudogap $\bar{\Delta}_{\rm PG}({\bf k})$ and SC gap $\bar{\Delta}_{\rm s}({\bf k}, \omega)$. However, this pseudogap has a largest value around the antinodal region, and then smoothly decreases upon increasing doping \cite{Feng15a,Feng12,Feng15}. This doping dependence of the pseudogap therefore leads to that the positions of the hump and low-energy peak around the antinodal region shown in Fig. \ref{PDH-doping} shift towards higher energies with the decrease of doping.

\section{Conclusions}\label{conclusions}

In conclusion, we have investigated the complicated line-shape in the electron quasiparticle excitation spectrum of cuprate superconductors based on the kinetic-energy driven SC mechanism. We show that the interaction between electrons by the exchange of spin excitations induces a well-pronounced peak structure in the electron quasiparticle scattering rate around the antinodal and nodal regions. However, this peak structure disappears at the hot spots, which leads to that the striking PDH structure in the electron quasiparticle excitation spectrum is developed around the antinodal and nodal regions, and then vanishes at the hot spots. The theory also shows that there is one to one correspondence between the peak structure in the electron quasiparticle scattering rate and the PDH structure in the electron quasiparticle excitation spectrum of cuprate superconductors, and all these unusual properties can be attributed to the emergence of the pseudogap.

\section*{Acknowledgements}

The authors would like to thank Dr. Yiqun Liu, Dr. Huaisong Zhao, and Professor Yongjun Wang for the helpful discussions. This work was supported by the National Key Research and Development Program of China under Grant No. 2016YFA0300304, and National Natural Science Foundation of China under Grant Nos. 11574032 and 11734002.

\end{document}